\documentclass[linenumbers]{aa}  

\usepackage{color}
\usepackage{chemformula}

\newcommand{\hi} {H{\sc i}}

\usepackage{graphicx}
\usepackage{txfonts}
\usepackage{amsmath}
\usepackage{multirow}
\usepackage{ulem}

\usepackage{natbib,twoopt}
\usepackage[breaklinks=true]{hyperref} %% to avoid \citeads line fills
\bibpunct{(}{)}{;}{a}{}{,}             %% natbib format for A&A and ApJ

\begin{document}

   \title{
   Secondary ionisation in hot atmospheres and interactions between planetary and stellar winds}

   \author{A. Gillet 
          \inst{1}
          \and
          A. Strugarek\inst{1}
          \and
          A. Garc\'ia Mu\~noz\inst{1}          
          }

   \institute{Université Paris Cité, Université Paris-Saclay, CEA, CNRS, AIM, F-91191, Gif-sur-Yvette, France\\
              \email{alexandre.gillet@cea.fr}
             }

 %  \date{Received September 15, 1996; accepted March 16, 1997}

  \abstract
   { The loss of close-in planetary atmospheres is influenced by various physical processes, such as photoionisation, which could potentially affect the atmosphere survivability on a secular timescale. The amount of stellar radiation converted into heat depends on the energy of the primary electrons produced by photoionisation and the local ionisation fraction. The {\hi} Lyman-alpha (Ly-$\alpha$ hereafter) line is an excellent probe for atmospheric escape but the origin of the high velocities detected in this line is not yet fully understood.
 }
{We characterise the effect of secondary ionisation by photoelectrons on the ionisation and heating of the gas for different planet-star systems in a 2D geometry. We study the interaction between the planetary and the stellar wind, the difference of the predicted mass-loss rates between 1D and 2D models, the signal of Ly-$\alpha$ and the impact of stellar flares.
}
   {
   Using the PLUTO code, we perform 2D hydrodynamics simulations for four different planets. We consider planets in the size range from Neptune to Jupiter. We produce synthetic Ly-$\alpha$ profiles to comprehend the origin of the signal and in particular its high velocity Doppler shift. 
   }
   {Our results indicate a trend similar to the 1D models, with a decrease in the planetary mass-loss rate for all systems when secondary ionisation is taken into account. The mass-loss rates are found to decrease by 48\% for the least massive planet when secondary ionisation is accounted for. We find nevertheless a decrease that is less pronounced in 2D than in 1D. We observe differences in the Ly-$\alpha$ profile between the different cases and significant asymmetries in all of them, especially for the lower mass planets. Finally, we observe that stellar flares do not affect the mass-loss rate because they act, in general, on a timescale that is too short.
   }
  %  {.}
  {We find and confirm in our extended 2D model that photoelectrons affect the mass-loss rates by factors that are potentially important for planetary evolution theories, and that they also affect the Ly-$\alpha$ profile. We find velocities in the escaping atmosphere up to 100 km/s, with the gas moving away from the star, which could be the result of the interaction with the stellar wind. We find that stellar flares generally occur on a timescale that is too short to have a visible impact on the mass-loss rate of the atmosphere.}

   \maketitle
%
%-------------------------------------------------------------------

\section{Introduction} \label{sec:intro}

Close-in exoplanets can be subject to a strong atmospheric escape due to the strong irradiation they receive \citep{lammer2003atmospheric}. The ionisation of upper planetary atmospheres and the heat deposition is thought to mostly originate from the absorption of the XUV (X-ray and extreme Ultra-Violet, from 15 {\AA} to 912 {\AA}) radiation from the host star \citep{yelle2004aeronomy,garciamunoz2007physical}. Depending on the ionisation conditions at the location where this flux is absorbed, the input energy can be primarily used to ionise, to heat, or to excite the atmosphere \citep{gillet2023self}. When the heating is sufficiently strong, it triggers a thermal atmospheric escape that is generally partially ionised (for a review of escape mechanisms, \citealt[see][]{gronoff2020atmospheric}). The escaping planetary material then interacts with the ambient, fully-ionised stellar wind, leading to further ionisation and neutralisation due to e.g. charge-exchange phenomena \citep{holmstrom2008energetic}. The extended tail of neutral, whether escaping neutrals from the planet or neutralised stellar wind, \citealt{tremblin2013colliding} leaves a detectable signal in transit of the Ly-$\alpha$ line around hot Jupiters \citep{vidal2003extended} and hot Neptunes \citep{kulow2014lyalpha}. In addition, strong escape can deplete rapidly the atmosphere of hot planets. Indeed, it could reduce the loss time of their atmospheres to a few hundreds of thousands of years for low-mass planets \citep{kubyshkina2018overcoming}, which can have a strong impact on their migration path as the stellar system evolves \citep{lazovik2023unravelling}. A precise, quantitative estimate of the atmospheric escape rate and of its detectable signals is therefore today strongly needed. 
\par
Numerous modelling efforts have been undertaken over the past decade to quantify the amount of planetary material susceptible to escape in this manner. Starting from 1D representation of the star-planet line, works such as \citet{yelle2004aeronomy} and \citet{garciamunoz2007physical} characterised the physical and chemical aeronomy of hot Jupiters, using the example of the hot Jupiter HD 209458 b. The modelling efforts continued, going from 1D \citealt{chadney2017effect,koskinen2022mass} to 3D \citep{rumenskikh2022global}, and considering various chemical compositions \citep{garciamunozetal2021,khodachenko2021impact}, multi-fluid approaches \citep{shaikhislamov2021global}, the effect of magnetic fields \citep{erkaev2017effect,daley2019hot}, the interaction with a magnetised stellar wind \citep{carolan2021effects} and NLTE effects \citep{garciamunozschneider2019}. 
\par
Despite these efforts, even in the simple case of an atmosphere composed purely of atomic hydrogen, the energy repartition between heating, excitation and ionisation was only recently quantified taking into account photo-electrons by \citet{guo2016influence} and \citet{gillet2023self} using 1D models of planetary atmospheres. They reported strong effects on the predicted mass-loss rate for hot Jupiters (to a reduction of a factor 2 of the predicted mass-loss rate when photoelectrons are taken into account), reinforcing the importance of well characterising the details of the energy deposition in the upper atmosphere. Conversely, the development of realistic planetary escape models should allow the community to directly compare model results with observed atmospheric escape tracers to put planetary escape theory on firm grounds, and leverage available observational information into characterisation of the aeronomy of planets.
\par
The repeated detection of Ly$\alpha$ transits in multiple systems also unveiled some temporal variability. For instance, \citet{des2012temporal} showed that the Ly$\alpha$ transit was visible for one epoch following a strong stellar flare of HD 189733, and not visible at another epoch, questioning if stellar variability could strongly influence the atmospheric escape. On one hand, \citet{chadney2017effect} showed with a 1D model of  HD 189733 b that the flare timescale was too short for an individual flare to have a significant effect on the atmospheric escape (but it could strongly influence the ionosphere of the planet). On the other hand, \citet{hazra2022impact} showed using a 3D model that a constant XUV flux corresponding to a strong flare should strongly impact the rate of atmospheric escape. As of today, the effect of successive, repeated strong flares on the atmospheric escape remains an open question. 
\par
In this work, we aim to extend the investigation carried out in  \citet{gillet2023self} by considering a more realistic 2D geometry, embedding a day and a night side for the atmosphere. A 2D geometry allows us to assess the effect of photoelectrons on the detectable absorption in Ly$\alpha$, while keeping the computational cost reasonable with respect to a 3D geometry. In our self-consistent treatment, we model the heating, ionisation and excitation caused by  XUV stellar photons in the atmosphere. The presentation of this work is organised as follows. We first present the 2D self-consistent model that we developed, as needed to describe
the thermal escape of the planetary wind and its interaction with the stellar wind (section \ref{sec:2}). We then present the comparison between the results obtained in 1D \citep{gillet2023self} and those obtained in 2D in section \ref{sec:3}. We describe our methodology to produce synthetic Ly-$\alpha$ transits from our simulations in section \ref{sec:4}, and show how photoelectrons and turbulent variability in our model affect such profiles in section \ref{sec:5}. We study the case of a strong flare in section \ref{sec:6}, and discuss and summarise our main results in section \ref{sec:7}.

\section{Model description}\label{sec:2}
\subsection{Extension of the 1D physical model}

Our 2D cartesian simulations are run with the hydrodynamics code PLUTO \citep{mignone2007pluto}, which solves the Euler equations in a rotating reference frame. Similarly to our 1D cases in \citet{gillet2023self}, the 2D equations for the conservation of mass, momentum, and energy solved in PLUTO are

\begin{equation}
\label{eq:1}
\frac{ \partial \rho}{ \partial t} + \boldsymbol{\nabla} \cdot (\rho {\bf u})= 0 \, ,
\end{equation}

\begin{equation}
\label{eq:2}
\frac{ \partial (\rho{\bf u})}{ \partial t} + \boldsymbol{\nabla} \cdot \left[\rho {\bf u} {\bf u} + P_T {\bf I}  \right] = - \rho \nabla \phi + \rho\bf({F}_{\rm cent} + {F}_{\rm cor})\, ,
\end{equation}

\begin{equation}
\label{eq:3}
\frac{ \partial (E + \rho\phi)}{ \partial t} + \boldsymbol{\nabla} \cdot [(E + P_T + \rho\phi){\bf u}] = \rho {\bf u} ({\bf F_{\rm cent} + {F}_{\rm cor}}) + H - C \, ,
\end{equation}

where $t$ is the time, ${\bf u}$ is the fluid velocity, and $\rho$ is the total density, $P_T$ is the thermal pressure, $E = P_T /(\gamma -1) + \rho {\bf u}^2/2$ is the total energy with $\gamma$ = 5/3, the adiabatic index for a mono-atomic gas. $C$ and $H$ are the cooling and heating terms defined in \citet{gillet2023self}. The joint gravitational potential of the planet plus the star is defined as
\begin{equation}
\label{eq:4}
    \phi = -\frac{GM_p}{r} - \frac{GM_\star}{R_{\rm orbit}-r}\,  ,
\end{equation}

where $G$ is the gravitational constant, and $M_\star$ and $M_p$ are the masses of the star and the planet, respectively.
The Coriolis and centrifugal force are defined as {\bf F}$_{\rm cor}$=$-$$2(\boldsymbol{\Omega} \times {\bf u}$) and 
{\bf F}$_{\rm cent}$= $\boldsymbol{\Omega} \times (\boldsymbol{\Omega} \times {\bf R})$ where ${\bf{R}}$ is the position vector relative to the star and $\boldsymbol{\Omega} = \sqrt{(GM_{\star}/R_{orb}^3)}$ {\bf e}$_z$ is the orbital rotation rate with ${\bf e}_z$ is the unit vector along the coordinate $z$. The model presented here solves these equations on the (x,y) orbital plane.
\par
We consider a planetary atmosphere composed of atomic hydrogen in neutral {\hi} and ionised H$^{+}$ forms plus thermal electrons. When the stellar radiation hits the planetary atmosphere, it ionises the gas and contributes to its heating. The photoionisation rate coefficient $J$ [s$^{-1}$] is given by:
\begin{equation}
\label{eq:J}
J = \int_{\lambda_{\rm min}}^{\lambda_0} \sigma_{\lambda} F_\star   \left(\frac{\lambda}{hc}\right) \left(1+\Phi_{\lambda,xe}\right) {\rm d}\lambda\, ,
\end{equation}
where $F_{\star}$ [erg cm$^{-2}$s$^{-1}${\AA}$^{-1}$] is the attenuated stellar flux. $\Phi_{\lambda,xe}$ is the number of secondary ions created per photoionisation. 
%\par
The heating deposition rate $H$ [erg s$^{-1}$ cm$^{-3}$] in the energy equation can be expressed as:
\begin{equation}
H = {n_\text{\hi}} \int_{}^{} \sigma_{\lambda} F_\star  \left(1-\frac{\lambda}{\lambda_0}\right) \eta_{\lambda,x_e}  {\rm d}\lambda\,
\label{eq:Hph}
\end{equation}
with $\eta_{\lambda,x_e}$ being the heating efficiency or rate of conversion from kinetic energy of the photoelectron into actual heating. The latter properly subtracts from the photoelectron energy the fraction of energy that goes into excitation and ionisation of the gas and, therefore, does not contribute to heating.
The full description of the chemistry and radiative transfer can be found in \citet{gillet2023self} along with the implementation of the photoelectrons. As there, the photoelectron physics is introduced through a parameterization inferred from calculations in \citet{garciamunoz2023a}.

\subsection{Stellar wind}\label{sec:stellarwind}

We consider the stellar wind as a fluid where all species (electrons, neutrals and ions) have the same velocity and temperature. The stellar wind is expected to have a significant impact on the structure and velocity of the planet's atmospheric escape. When the planetary wind meets the stellar wind, its structure changes in a way similar to that of a comet. Structures of different sizes are then formed: bow shock, turbulence as well as instabilities (such as shearing due to the velocity differential between the two winds). The overall structure of the planetary wind can vary depending on the ram and thermal pressure ratios of each wind, and have been classified in particular by \citet{matsakos2015classification}. 
\par

\begin{figure*}[ht!]
\centerline{\includegraphics[width=\linewidth]{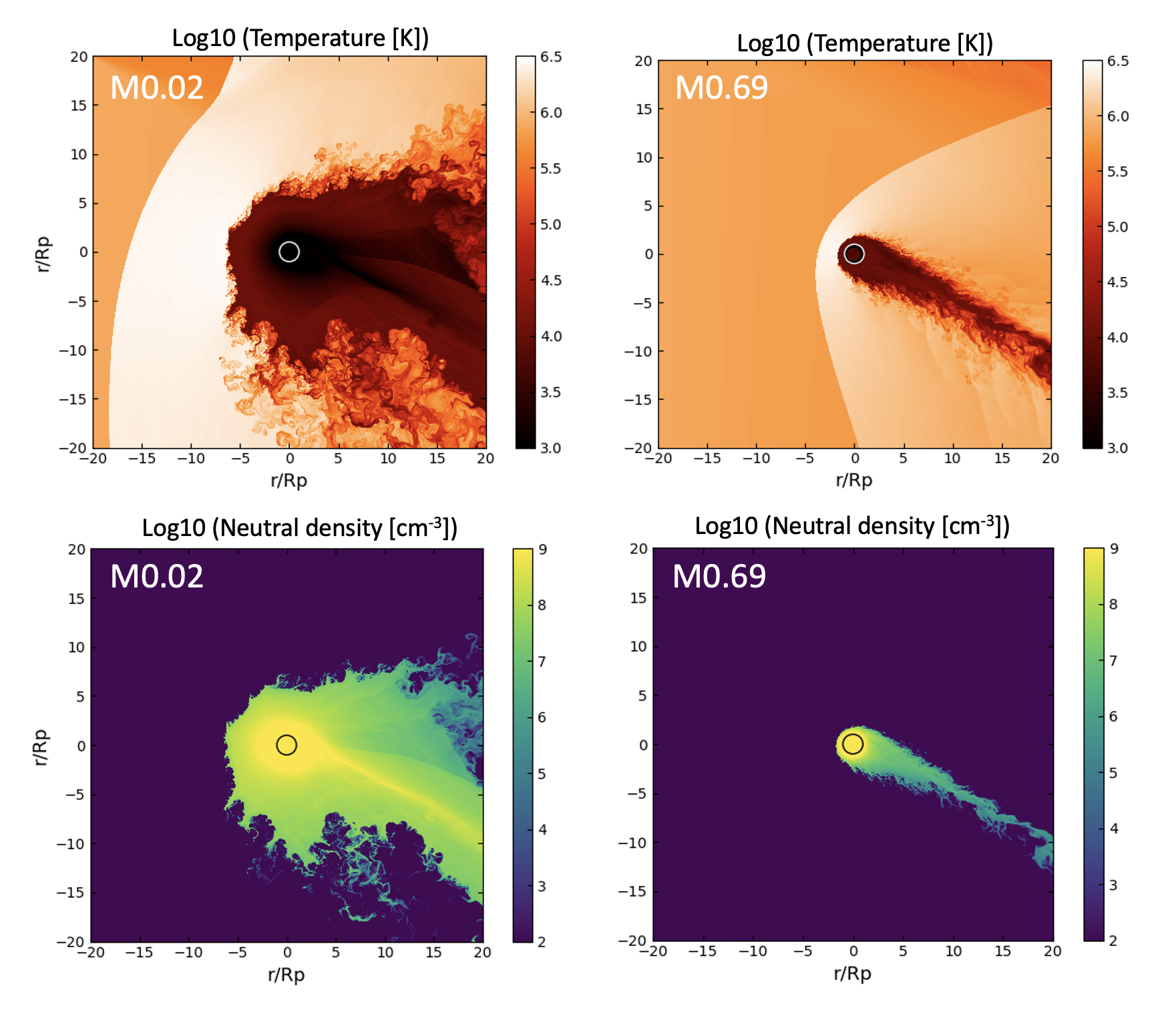}}
\caption{Illustration of the dynamic characteristics of the interaction of the planetary wind (with the planet at the center delimited by a circle) and stellar wind for the most extreme cases M0.02 (left) and M0.69 (right). \textit{Top panel}: Instantaneous temperature in Kelvin in logarithmic scale. \textit{Bottom panel}: Instantaneous density of neutrals in logarithmic scale.}
\label{fig:Temp2D}
\end{figure*}

If the stellar wind is supersonic, it will form a shock as it slows down to adapt to the velocity of the obstacle (the planet in this case). The velocity of the post-shock stellar wind seen by the planet will therefore be lower than the pre-shock velocity (via the Rankine-Hugoniot shock conditions). The interaction of the two winds results in an equilibrium point of thermal pressure and ram pressure (see figure \ref{fig:ramthermal}) with $P_{ram} = \frac{1 }{2}\rho v^2$. 
If the equilibrium point is achieved between the ram pressure of the planetary wind and the thermal pressure of the stellar wind, then the mass-loss rate of the planetary escape remains independent of the stellar wind conditions \citep{christie2016axisymmetric}. If the equilibrium point is achieved by the thermal pressures, then we have a modification of the boundary conditions of the planetary escape. Strong stellar winds can prevent the planetary wind from reaching supersonic speeds in the vicinity of the planet \citep{vidotto2020stellar,carolan2021effects}. It is therefore crucial to properly model the stellar wind in order to study these interactions with the planetary wind.
\par
In our model, the stellar wind is injected into the 2D Cartesian grid at the left edge of the domain along the $y$ axis. We based our stellar wind prescription on the wind model of \citet{parker1958dynamics}. We impose a stellar wind mass-loss rate as a fixed parameter in our model and set it to 50 $\dot{M_\odot}$ with $\dot{M_\odot}$ = 2 $\times$ $10^{-14}$ $M_\odot$ $yr^{-1}$ (see also \citealt{carolan2021effects}). Such a large mass-loss is representative of young Sun-like stars \citep{wood2021new} and leads to a stellar/planetary wind interaction located within 20 planetary radii, which appropriately limits the size of the box required for our modeling. Indeed, a weaker stellar wind does not allow us to contain the planetary wind in a reasonable numerical box size. For close-in planets, a global modelling embedding also the star itself is then needed \citep[see][]{matsakos2015classification,strugarek2015magnetic}. The stellar wind temperature is set to T$_{w}$ = 1.5 $\times$ 10$^{6}$ K. The density $\rho$ is prescribed by the stellar mass-loss as $\rho_w$ = $M_\odot$/4$\pi$r$^2$v$_{w}$ with v$_{w}$ = 400 km/s and the pressure can be derived from density and temperature. We correct v$_x$ and v$_y$ to take into account the rotating frame. The neutral fraction x$_{HI}$ is set to zero when the stellar flow enters the box.

\subsection{Numerical model}

We solve the Euler equations on a 2D cartesian grid,x, y = [-20, 20]$R_{p}$ parallel to the orbital plane with the planet located at center of the box. We carry out high resolution 2D simulations (5000 cells $\times$ 5000 cells), cut into a uniform grid of 1000 $\times$ 1000 cells covering the square [-1,1] around the planet, and stretched by 2000 points in each direction between $\pm 1 $ and $\pm$20 $R_{p}$. In all our cases, we initialise our model by interpolating the corresponding 1D spherical converged solution (obtained with the model described in \citealt{gillet2023self}) on the 2D Cartesian grid. We aimed to have the highest possible resolution between the surface of the planet and the interface of interaction with the stellar wind in order to study the mechanisms of photoionisation, energy deposition, but also resolve the small structures of the planetary wind.
\par
For the numerical method, we use an advanced option in PLUTO named "Shock flattening" in its "MULTID" version \citep{colella1984piecewise} which provides more dissipation in the star-planet interaction region where the shock occurs. To solve the Euler equations, as in 1D, we use the Riemann solver in the Harten-Lax-Van approximation, resolving exactly the contact discontinuities between the cells \citep[HLLC, see][]{toro2009hll}. We use a linear reconstruction for the spatial order of integration and a third-order Runge-Kutta (RK3) scheme for the temporal evolution. Typically, a run takes about two weeks to reach a relaxed state on 400 cores.
\par
In the same way as in the 1D case \citep{gillet2023self}, at the interior boundary of each planet, we fixed the density to 1.326 $\times$ 10$^{-10}$ g/cm$^{3}$, the pressure to 12 $\mu$bar and the velocity to zero, and we assumed that the gas is entirely neutral. This pressure is typical of 1D models and ensures that at the base of the model atmosphere the gas remains in hydrostatic equilibrium and most of the stellar XUV energy is deposited at higher altitudes. The local temperature at the level of the atmosphere is 1100 K, which is broadly representative of conditions in close-in planets. We prescribe the same boundary conditions at the interior limit of each planet. The planet is now an internal boundary condition (CL) in our grid. The circular limit of this CL is therefore discretised on the cartesian grid. We have described above how the edge through which the stellar wind enters is treated. Next, we describe the other three edges : the condition at the outer boundaries of the box on the other three edges is defined by a zero gradient for $\rho$, P and x$_{HI}$ in all circumstances. In the case of velocity, we prescribe a zero gradient if the flow leaves the box and set it to a zero value if it enters the box at any of the three boundaries where the stellar wind is not imposed. 
\par

In this study, we cover the same planetary masses as in \citet{gillet2023self}: 0.02 0.05, 0.1, and 0.69 $M_{J}$ (see Table \ref{table:1}). They range from a sub-Neptune planet to a Jovian-like planet similar to HD209458b. 

\begin{table}[ht]
\centering
    \caption{Planet parameters for each planetary system M0.69-M0.02.}
 %\begin{tabular}{|c|c|c|c|c|}
 \begin{tabular}{ccccc}
  \hline\hline
  Parameters&M0.69&M0.1&M0.05&M0.02\\
  \hline
  $M_{p}$ ($M_{\rm J}$) & 0.69 & 0.10 & 0.05 & 0.02\\
  $R_{p}$ ($R_{\rm J}$) & 1.32 & 0.69 &  0.55 & 0.40\\
 \hline

 \end{tabular}%}
\tablefoot{$M_{p}$ and  $R_{p}$ are the planetary mass and radius respectively in Jovian units. } 
 
 \label{table:1}
\end{table}

For the stellar XUV irradiation, we adopted a solar spectrum downloaded from the SOLID data exploitation  project,\footnote{European comprehensive solar irradiance data exploitation; https://projects.pmodwrc.ch/solid} as observed on December 13, 2021. We used $R_\star$ = 7 $\times$ 10$^{10}$ cm and  $M_\star$ = 1.98 $\times$ 10$^{30}$ g as stellar radius and mass respectively and an orbital distance of 0.045 AU. The XUV-integrated flux of the star at 1 AU is 4.4 erg cm$^{-2}$ s$^{-1}$, and it is about 500 times larger at the orbit of the planet. We assume the planet is tidally locked and spins and orbits with the same period.

\section{Self-consistent hydrodynamic escape in 2D}\label{sec:3}

\subsection{Dynamics of the planetary outflow}

The planetary wind forms due to the interaction of incident XUV radiation from the host star with all layers of the planet's atmosphere. It produces an extended envelope of partially ionised gas that escapes and is directed against the comparatively hot, fast, fully ionised stellar wind.
\par
In our simulations, we capture the formation of multi-scale vortices for all our cases. These simulated instabilities only appear at high resolution \citep{tremblin2013colliding}. In all cases, the bow shock is created in front of the planet upward to the interaction of the two winds (stellar + planetary) whichever the resolution. These dynamic characteristics are illustrated in figure \ref{fig:Temp2D}, for the M0.02 case on the left panels and for the M0.69 case on the right panels.
\par
We observe that the temperature (top panels) can rise up to several million kelvins in the zone between the bow shock and the planet's atmosphere, creating an extreme temperature gradient. The atmosphere of the least massive case extends up to several planetary radii in the line of sight of the star on the day side, with the creation of vortices lasting beyond 15 $R_{p}$ in the cometary tail of the planet. Case M0.69 has a highly compressed atmosphere with a smoother planetary tail seemingly having fewer vortices. In case M0.02, neutrals can be found up to 5 $R_{p}$ in all directions (see lower panels of figure \ref{fig:Temp2D}) whereas they are limited to a region much closer to the planet (in relative scales) in case M0.69. In addition, case M0.02 displays a more complex structure with a denser region of neutrals inside the tail.

\begin{figure}[h!]
\centerline{\includegraphics[width=\linewidth]{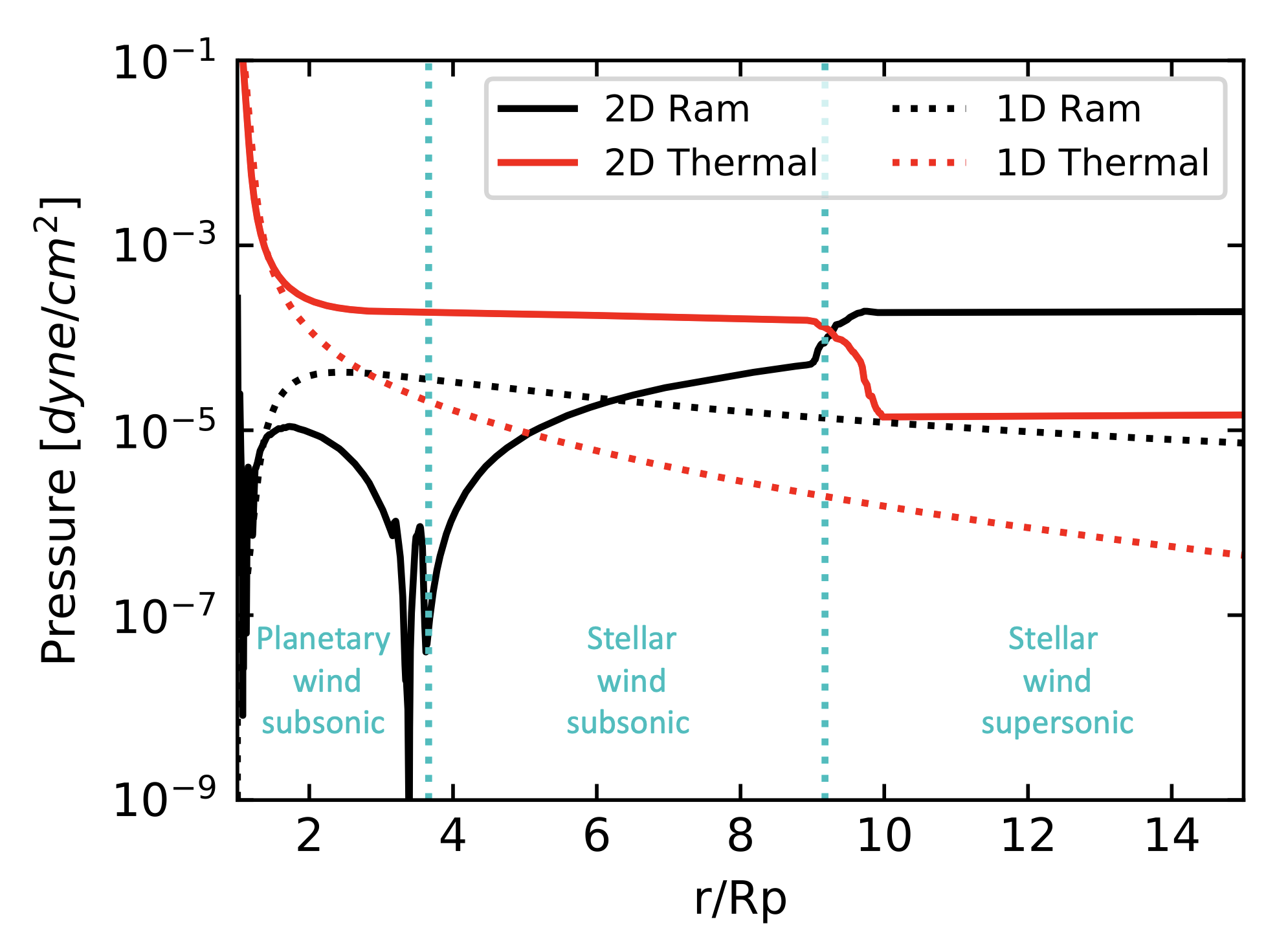}}
\caption{Thermal and ram pressure profiles for the M0.05 case. Comparison between 1D (dashed lines) and 2D (solid lines) simulations. In 2D, the profiles are taken on the star-planet line on the day side and averaged over a time period of 42 h.}
\label{fig:ramthermal}
\end{figure}

\begin{figure}[h!]
\centerline{\includegraphics[width=\linewidth]{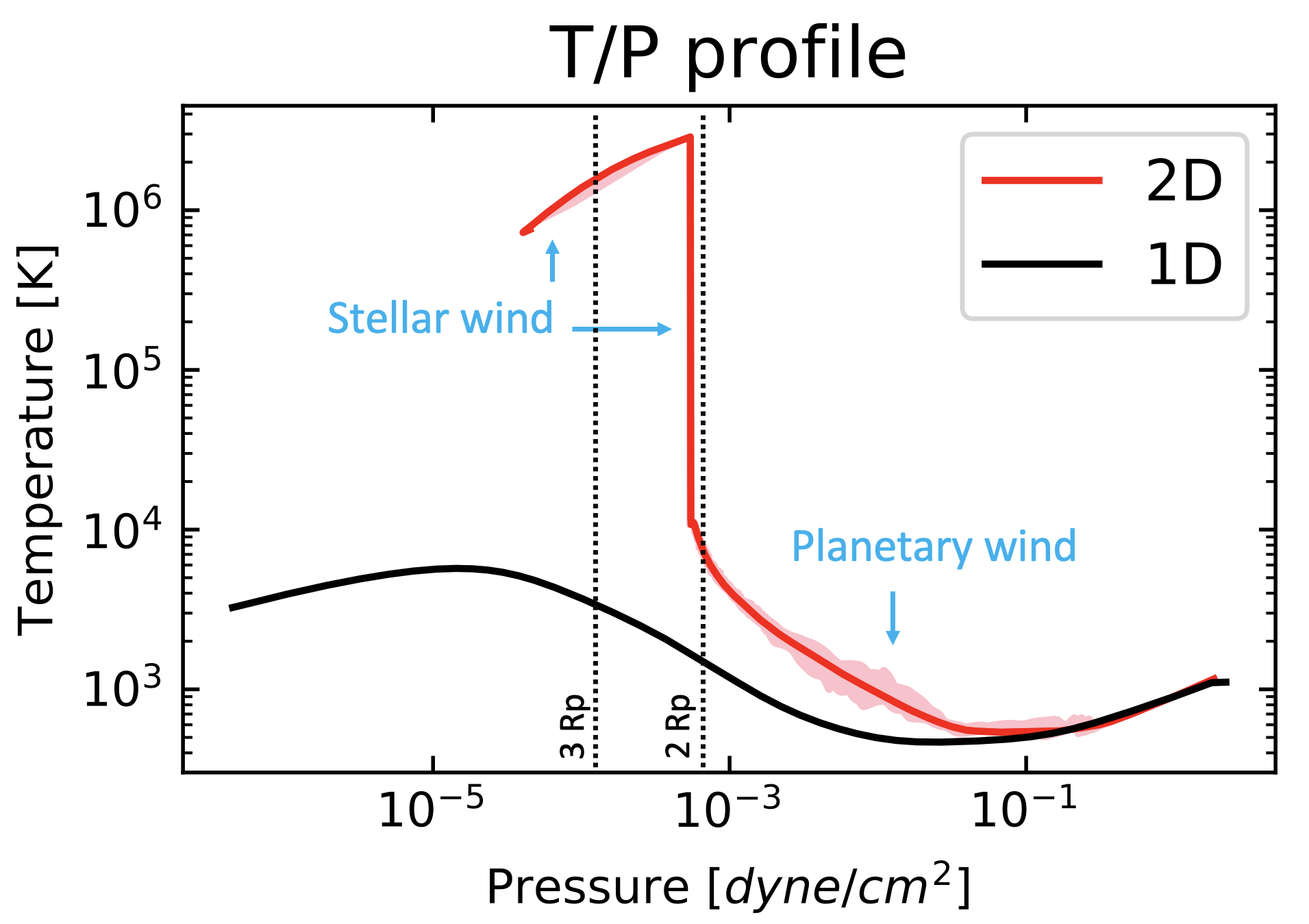}}
\caption{Pressure/Temperature profile for the M0.05 case. Comparison between the 1D profile, in black, with the 2D profile in red, averaged over 42 hours. Temporal variability is shown in pink averaged over half of the orbital period. 1 $R_{p}$ correspond to the surface at which we start studying the upper atmosphere.}
\label{fig:T_P}
\end{figure}

The ram pressure at the sonic point of the stellar wind is one of the essential parameters which determines whether or not the exoplanetary atmosphere is confined by the stellar wind \citep{vidotto2020stellar}. Indeed, if the ram pressure of the stellar wind is greater than the total pressure of the planetary wind (their ratio is equal to ${M}^2$ with $M$ the Mach number), the latter  remains subsonic and is therefore sensitive to the confinement by the stellar wind. In the case of our simulations, all our atmospheres are confined. We illustrate this in figure \ref{fig:ramthermal} where we can see that below the interface between the subsonic and supersonic stellar wind, the total pressure ($P_{ram} + P_{T} $) of the prescribed stellar wind is greater (above 4 $R_{p}$) than that of the planetary wind (below 4 $R_{p}$). For the M0.05 case, deep in the atmosphere, the ram and thermal pressures of the planetary wind in 1D and 2D overlap up to 1.5 $R_{p}$ (solid and dashed lines), these then differentiate, until completely diverging at the interface between the subsonic planetary wind and the subsonic stellar wind at 2.8 $R_{p}$. 
\\
Deep in the atmosphere, the pressure and temperature in the 2D simulations correspond to the equivalent 1D case \citep{gillet2023self} as shown by the temperature-pressure profile in figure \ref{fig:T_P}. However, in the upper atmosphere, the 2D case deviates considerably from the 1D case, as expected, due to the deviation from the 1D steady state (represented by the black curve) and modulated by the interaction of the two winds leading to a compression of the planetary atmosphere. We note the small temporal variability on the T/P profiles vs r/$R_{p}$ of all our cases, calculated over a time period around 42 hours corresponding to about half of the 3.5 days orbital period of the planet(pink area, in figure \ref{fig:T_P} for the case M0.05).

\subsection{Secondary ionisation by photoelectrons in 2D}

We now focus on the effect of photoelectrons, and compare the photoionisation and heating rates from our 
2D simulations with those from the 1D 
geometry \citep{gillet2023self}. In this section, we illustrate it with a mini-Neptune type planet (case M0.05 in table \ref{table:1}).

\begin{figure}[h!]
\centerline{\includegraphics[width=\linewidth]{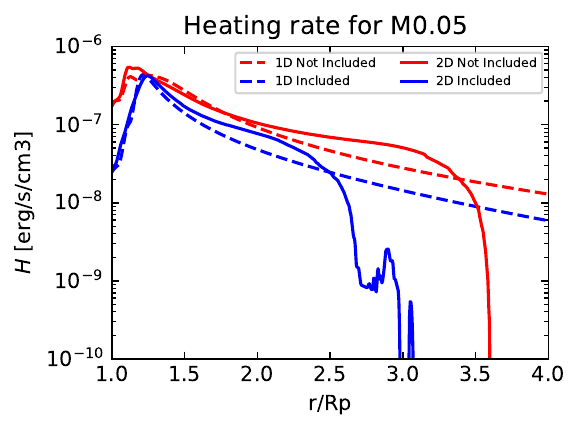}}
\centerline{\includegraphics[width=0.96\linewidth]{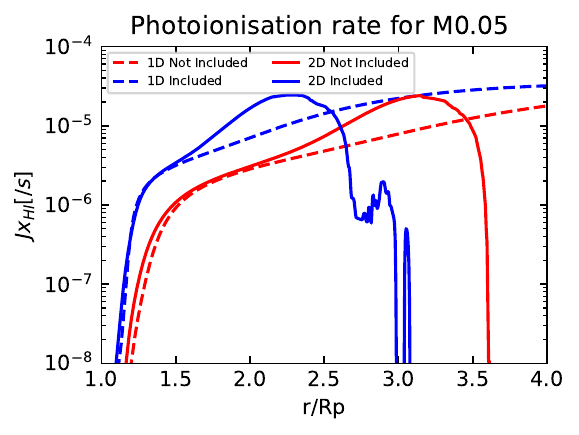}}

\caption{Illustration of the heating rate along with the photoionisation rate and its dependence on the ionisation fraction $X_{HI}$ as a function of the radial distance on the day side and averaged over 42 hours. \textit{Top panel}: comparison of the heating rate for the M0.05 case in 1D and 2D, with (in blue) and without (in red) the photoelectrons. \textit{Bottom panel}: comparison of the photoionisation rate J$\times$$x_{HI}$ for the M0.05 case in 1D and 2D, with (in blue) and without (in red) the photoelectrons.}
\label{fig:1D2DJ_H}
\end{figure}

We first describe the case where secondary ionisation is not included (red curves) in the top panel of figure \ref{fig:1D2DJ_H}. In this approach,  all the excess energy above the ionisation threshold goes into heating.
We observe that, in both the 1D (dotted lines) and 2D (solid lines) cases, the heating rates peak at similar altitudes. Both the 1D and 2D solutions have two maxima. The two maxima are more easily discernible in the 1D case. The 2D case decreases more slowly in the upper atmosphere up to 3.5 $R_{p}$, this is because the planetary wind is comparatively denser, as seen in figure \ref{fig:T_P}. Above this altitude, the ionisation fraction is close to 1, because the fully ionised stellar wind dominates, and the heating deposition rate naturally drops sharply. At low altitude, the photoionisation rates of both 1D and 2D cases increase similarly, up to 1.3 $R_{p}$. 

At higher altitudes, the photoionisation rates of the 2D case (bottom panel) increase more rapidly, as a result of the compression exercised by the stellar wind. When secondary ionisation is taken into account (shown by the solid and dashed blue lines in figure \ref{fig:1D2DJ_H}), we observe a shape similar to the thermal deposition rate curve in the two cases. The ionisation rate nevertheless increases more quickly in 2D between 1.5 $R_{p}$ and 2.4 $R_{p}$ before dropping due to the interaction with the stellar wind. 
\par
However, we observe significant differences in 2D when secondary ionisation is included (blue) or not (red). The collision point of the two subsonic flows is located at 2.6 $R_{p}$ for the case of photoelectrons, while it is located much further up at 3.5 $R_{p}$ when these are not included. The planetary wind is shaped by the balance between its thermal pressure and the ram pressure of the stellar wind. The total pressure of the case with photoelectrons being lower, due to smaller loss rates and therefore velocities, the stellar wind compresses the planetary wind more. The 2D profile for the photoionisation rate deviates between 1.5 and 2.6 $R_{p}$ (peak in the bottom panel of figure \ref {fig:1D2DJ_H}) with a stronger rate in the 2D geometry due to a greater quantity of neutrals at this altitude. Finally, at the lowest altitudes, the geometry has very little impact on the heating and ionisation rates on the star-planet axis.
\par

\begin{figure}[h!]
\centerline{\includegraphics[width=\linewidth]{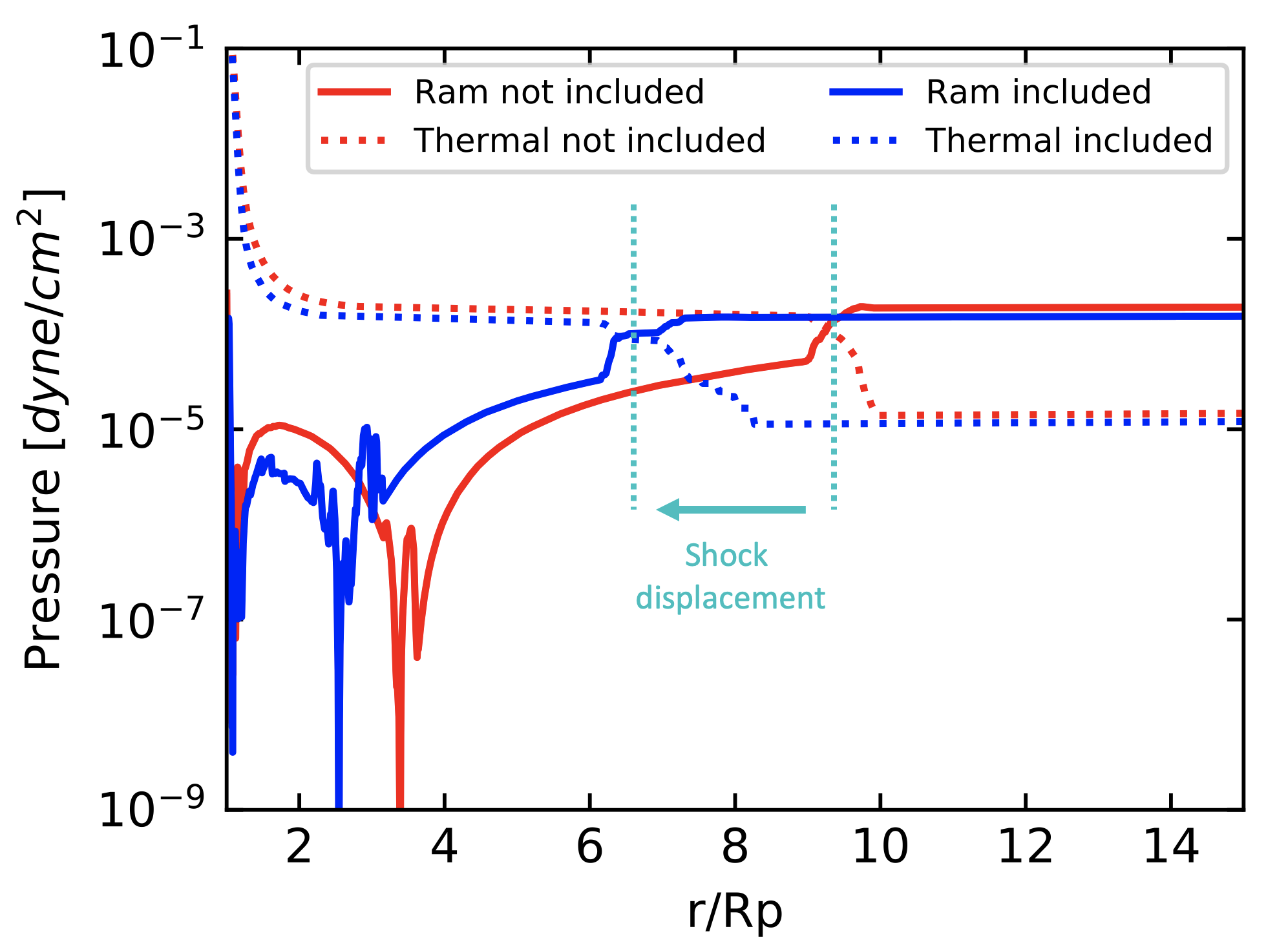}}
\caption{Ram (solid line) and thermal (dashed line) pressure profiles for case M0.05 with photoelectrons included or not, averaged over 42 hours. Representation of the effect of secondary ionisation without (in red) and with (in blue) on the pressures of planetary and stellar winds.}
\label{fig:ram_thermal_photoelec}
\end{figure}

The ram (in red) and thermal (in blue) pressure profiles are also affected, as shown in figure \ref{fig:ram_thermal_photoelec}. The planetary ram pressure shows different structures when photoelectrons are included (dashed lines), while the thermal pressure decreases. 
The stellar wind shock moves closer to the planet when photoelectrons are considered, influencing the ram pressure structure deep in the atmosphere.

\subsection{Atmospheric mass-loss rates}

\begin{figure}[h]
\centerline{\includegraphics[width=1.0\linewidth]{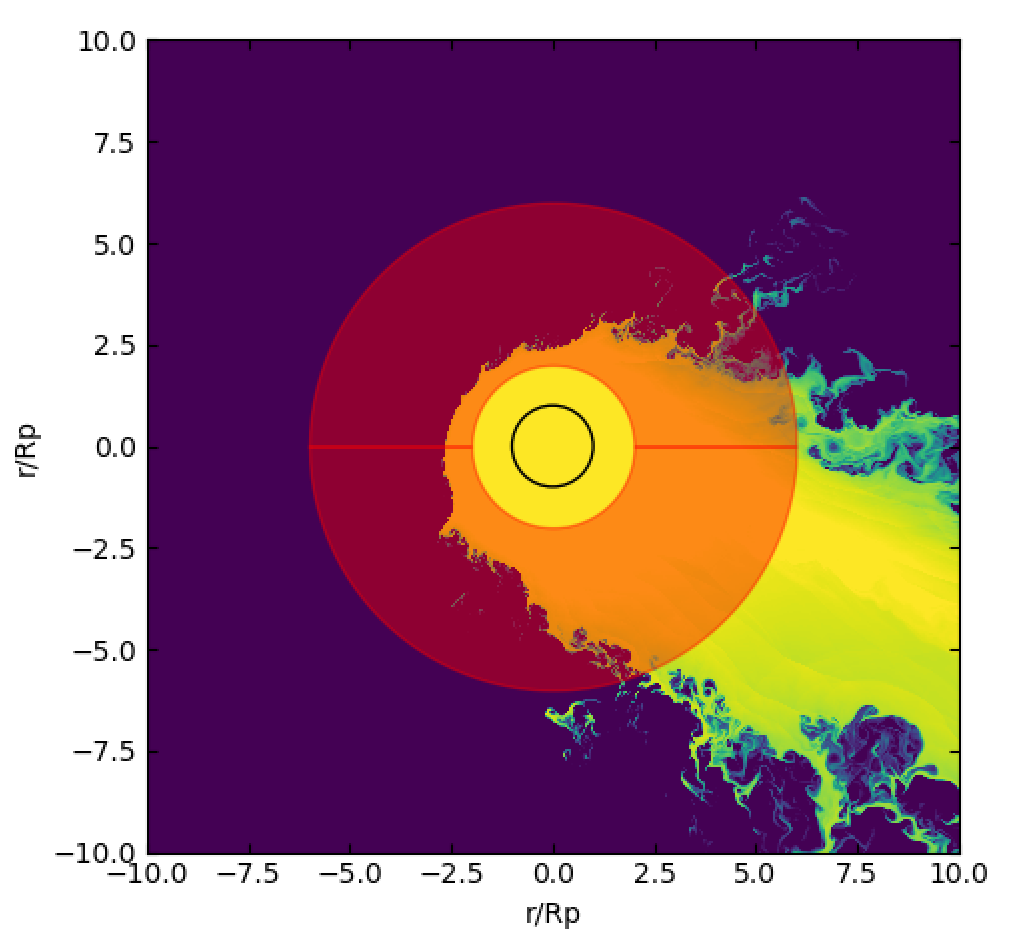}}
\caption{Illustration of the calculation of the mass-loss rate in 2D for the M0.1 case. The mass flux integration ring is represented by the red ring.}
\label{fig:massloss_illu}
\end{figure}

In the 2D geometry we are concerned with, the closed area over which the flux of mass must be calculated to determine the mass loss rate is the area of a cylinder.
In a first step, the flux of mass is calculated through an annulus around the planet of radius, depending on the case, between 2 and 6 $R_{p}$. This range helps stabilise the mass-loss rate, meaning that it minimizes the temporal variations in and out of the integration volume. One example is shown in figure \ref{fig:massloss_illu}. In a second step, 
we multiply the above determination by $L$=$2R_p$, where $L$ represents the height of the cylinder. This choice ensures the match between the expressions for the mass loss rate between the 1D (spherical shell symmetry) and our current 2D geometry. We note that because the 1D and 2D geometries differ in their radial extension, the mass loss rates are not expected to match exactly between the two types of models even when using $L=2R_p$.
We report in table \ref{table:2D} the mass-loss rate in 2D for all cases.

\begin{table}[ht]
\centering
   \caption{Mass-loss rates
}
\begin{tabular}{cccc}
 \hline\hline
\multicolumn{1}{l}{\multirow{2}{*}{Planet}} & \multicolumn{1}{l}{Photoelectrons} & \multicolumn{1}{l}{Photoelectrons} & \multicolumn{1}{l}{\multirow{2}{*}{Diminution}} \\
\multicolumn{1}{l}{}                        & not included                       & included                           & \multicolumn{1}{l}{}                            \\ \hline
 M0.69 & 2.31 & 1.78 &22\%\\ 
 M0.1   & 5.16 & 3.43 &33\%\\
 M0.05  & 4.96 & 2.79 &43\%\\
 M0.02  & 5.36 & 2.80 &48\%\\ 
\hline
\end{tabular}
\tablefoot{Atmospheric mass-loss rates [$10^{10}$ g/s] calculated in 2D for each of the exoplanetary systems in this work, with photoelectrons included or not.} 

\label{table:2D}
\end{table}

We observe a reduction in mass-loss rate from 22$\%$ (M0.69) to 48$\%$ (M0.02) when considering the effect of photoelectrons in the 2D cases. The same dependence on planetary mass as in 1D is shown: secondary ionisation impacts low mass planets more easily. The variation in mass-loss is slightly less pronounced in 2D than in 1D, for example a reduction of 48\% in 2D, compared to 54\% in 1D for the M0.02 case.

\begin{table}[ht]
\centering
   \caption{Differences between 1D and 2D}
\begin{tabular}{ccc}
 \hline\hline
\multicolumn{1}{l}{\multirow{3}{*}{Planet}} & \multicolumn{2}{c}{2D/1D}                                               \\ \cline{2-3} 
\multicolumn{1}{l}{}                        & \multicolumn{1}{l}{Photoelectrons} & \multicolumn{1}{l}{Photoelectrons} \\
\multicolumn{1}{l}{}                        & not included                       & included                           \\ \hline
 M0.69 & 91\% & 88\%\\ 
 M0.1  & 81\% & 76\%\\
 M0.05 &82\%&  80\%\\
 M0.02  & 83\% & 81\% \\ 
\hline
\end{tabular}
\tablefoot{Relative difference in atmospheric mass-loss between 2D and 1D with photoelectrons included or not. 100\% corresponds to the same value between 2D and 1D, a lower percentage indicates a decrease in 2D.}
\label{table:1D2D}
\end{table}

We also observe a difference between the mass-loss rate in 1D and 2D, as reported in table \ref{table:1D2D}. The relative 1D-2D difference is reported as a percentage. In all cases, the mass-loss rate is slightly lower in 2D than in 1D. For example, the 2D M0.69 case presents a mass-loss rate of 91\% of the analogous 1D case without photoelectrons. This is probably due that, in 2D, only the day part of the atmosphere receives the incident flow. In addition, we have shown previously, the planetary wind is confined and compressed in our 2D models as we have shown previously, which a priori leads to a moderate reduction in the mass-loss rate, as predicted in \citealt{vidotto2020stellar}.
\par
We have run one more simulation with the 2D geometry and without stellar wind including the specifics of photoelectrons in order to assess the 1D-vs-2D effects. We obtained a mass-loss rate of 5.0 $\times 10^{10}$ g/s, that should be compared with the 1.78 $\times 10^{10}$ g/s for the case with stellar wind. This shows that, in the 2D case, stellar wind confinement reduces the mass-loss rate by a factor of 3.

\section{Ly-$\alpha$ absorption of planetary atmospheric escape}\label{sec:4}

\subsection{Theory of Ly-$\alpha$}

The Ly-$\alpha$ absorption profile can provide insight into the magnitude of the atmospheric escape process \citep{2023MNRAS.518.4357O}. The relevant information is encoded in the signal strength and its wavelength dependence. We use our simulation results to calculate synthetic transmission spectra. The absorption cross-section of a line is:

\begin{equation}
\label{eq:5}
\sigma(\lambda) = \frac{\pi e^2}{m_e c^2} f_{lu} \lambda_0^2 {\psi} (\lambda, u, T),
\end{equation}

where $e$ is the electronic charge, $m_e$ the electronic mass, $c$ the speed of light, $\lambda_0$ is the wavelength of the line at rest equal to 1215.67 {\AA} and $f_ {lu}$ is the strength of its oscillator 0.41641 for the Ly-$\alpha$ line \citep{kramida2018nist}.
\par
The so-called Voigt profile ${\psi}(\lambda, u, T)$ (normalised so that $\int{\psi}(\lambda, u, T)d\lambda=1$) is a convolution of a Lorentzian profile resulting from the natural widening (which dominates in the wings) and a Gaussian profile resulting from the Doppler shift, dominating in the centre of the line. It is calculated by Doppler effect ($\lambda$-$\lambda_0$)$(1-{u}/c)$ where $u$ is the velocity component along a single ray in the star-planet line of sight.
\par
The Voigt distribution is calculated with width at half maximum for the Doppler component $\Delta \lambda_D = \lambda_0$ $\times$ 7.16 $\times$10$^{-7}$ $\sqrt {T}$, with T the local temperature in Kelvin and the width at half maximum for the Lorentzian component $\Delta \lambda_L$ = $\lambda_0^2$ $A_{ul}$ / (2$\pi c$), where $A_ {ul}$=6.2648 $\times$ 10$^{8}$ s$^{-1}$ is the transition probability of the line \citep{kramida2018nist}.
\par
The differential optical thickness at wavelength $\lambda$ for a ray passing through a grid element along the line of sight from the star to the planet (for all x component in the box) is calculated by:

\begin{equation}
\label{eq:6}
\Delta \tau(\lambda,x,y) = n_{\text{\hi}}(x,y) \;  \sigma(\lambda) \; dx,
\end{equation} 

where n$_{\text{\hi}}$ is the neutral particle density of hydrogen and $\sigma(\lambda)$ is the absorption cross-section. The optical thickness $\tau_{\lambda}$($y$) at a given position in the $y$ direction in our 2D spatial grid is calculated as follows:

\begin{equation}
\label{eq:7}
\tau_{\lambda}(y) = \sum_{x} \Delta \tau(\lambda,x,y).
\end{equation}

Finally, we can calculate an effective size of the planet for a given wavelength $\lambda$ in our 2D geometry as follows:

\begin{equation}
\label{eq:8}
 D_{\lambda} =  \int_{-\infty}^{+\infty} \left[ 1 - \exp{(-\tau_{\lambda}(y)})  \right] dy.
\end{equation}

To construct the synthetic Ly-$\alpha$ profile, we choose 100 wavelength intervals over the wavelength range [1214.17-1217.17 {\AA}]. We verified that this choice did not impact the results we present by increasing the spectral resolution by a factor of two on a few selected tests. We then calculate a wavelength-dependent effective radius as R$_{\rm eff}$ = $D_{\lambda}$/2 and compare it to the solid optical radius R$_{\rm opt}$ corresponding to a planet without atmosphere. We then calculate a wavelength-dependent effective radius as R$_{\rm eff}$ = $D_{\lambda}$/2 and compare it to the solid optical radius R$_{\rm opt}$ corresponding to a planet without atmosphere (what is measured with optical photometry).

\subsection{Origin of the signal for a hot Jupiter}

To better understand the origin of the Ly-$\alpha$ signal, we illustrate the atmosphere of the M0.69 case. As shown in equation \ref{eq:6}, the signal is sensitive to the particle density of neutral hydrogen n$_{\text{\hi}}$, the temperature and the velocity of the gas (both contained in ${\psi}(\lambda, u, T)$). We expect a strong signal contribution near the planet, where neutral densities are high, and no contribution in regions where the fully ionised stellar wind dominates. As for the shape of the absorption line, it will be partly dictated by the structure of the neutral gas flow velocity.

\begin{figure}[h]
\centerline{\includegraphics[width=\linewidth]{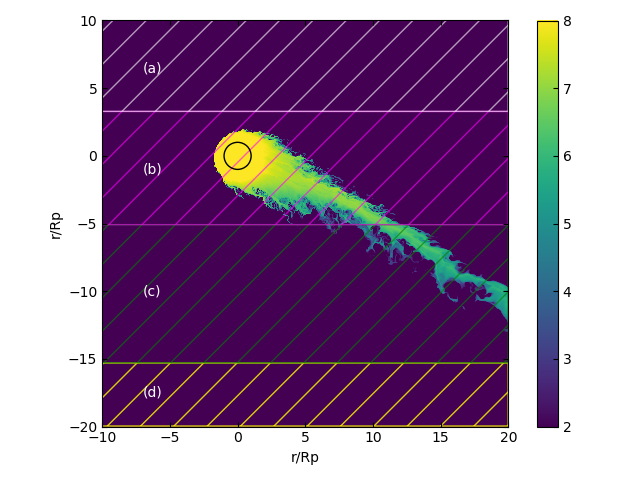}}
\centerline{\includegraphics[width=\linewidth]{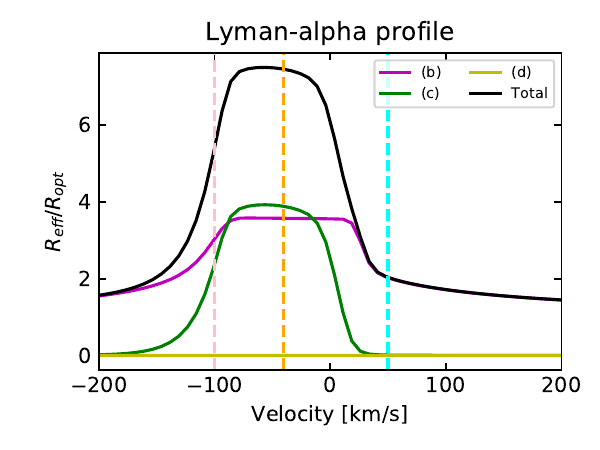}}
\caption{Illustration of the Ly-$\alpha$ signal. \textit{Top panel}: 2D representation of the instantaneous number density of gas escaping from a Jupiter-like planet. The coloured hatching shows different regions scanned in our simulation grid to obtain the Ly-$\alpha$ profile. \textit{Lower panel}: Results of instantaneous Ly-$\alpha$ profiles associated with the different regions scanned around the planet.}
\label{fig:Lymanillustration}
\end{figure}

We split the signal R$_{\rm eff}$/R$_{\rm opt}$ into several contributions in the $y$ direction near the planet (see the top panel of Fig \ref{fig:Lymanillustration}). As expected, we observe (bottom panel) that two regions contribute the most to the Ly-$\alpha$ signal, those in magenta (b) and green (c) colours. In the same panel, the total signal is represented by the black line. Indeed, the two regions scan the planetary atmosphere where the particle density of neutral hydrogen is the highest as well as the planetary tail where we can still find neutrals up to 20 $R_{p}$ in the positive $x$ direction.
\par
The contribution of zone (b) is saturated at the core of the line, contributing to the total signal by the geometric size of the scanned region. As expected, there is a very small contribution from region (a) and region (d), with less than 0.01$\%$ of the total signal, both regions combined. Ionised stellar wind dominates in these regions, preventing the build-up of a significant Ly-$\alpha$ signal. On the other hand, the wings of the distribution are dominated by the purple part (b) of the domain due to a greater density of neutrals and the temperature of the gas.
\par
\begin{figure}[h]
%\centerline{\includegraphics[width=\linewidth]{velocityM0_05.png}}
\centerline{\includegraphics[width=\linewidth]{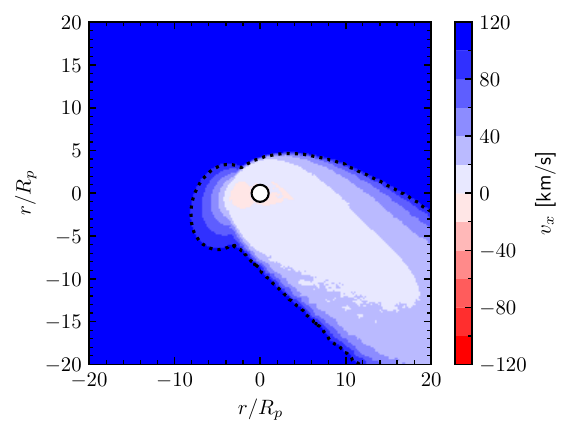}}
\caption{Illustration of the 42h-averaged velocity v$_x$ profile computed at the inertial frame for the case M0.05 along with a velocity line of 100 km/s showed by the dotted black contour.}
\label{fig:velocities}
\end{figure}

Figure \ref{fig:velocities} shows the velocity v$_x$ of the flow. The material escapes with relatively modest velocities of 20-40 km/s. However, at the interface between the planetary and stellar wind in the planet's wake, the velocities reach  50 to 100 km/s (showed by dotted black contour in figure \ref{fig:velocities}).  This acceleration is  caused by the interaction with the stellar wind. Indeed, to assess why such velocities are reached, we went back to the extra model that we ran without stellar wind. This model (not shown here) presents absorption at large velocities up to $|v|\simeq$ 100 km/s, but without a clearly preferred blue or red shift. The large blueshift seen in the figures shown in the manuscript therefore  originates from the interaction of the planetary wind with the stellar wind, which also helps reach higher speeds closer to the planet due to entrainment.

The peaks of the two dominant contributions (regions (b) and (c) in the upper panel of figure \ref{fig:Lymanillustration} are all blue-shifted due to the stellar wind accelerating the gas to high velocities towards the planetary tail and hence towards the observer. The total signal is blue-shifted, resulting in a particularly noticeable contribution at a speed of 100 km/s on the left wing of the line. Unlike the 1D case where we observe lower speeds, here, the speeds observed in 2D are caused by the entrainment of neutrals by the stellar wind. The asymmetry of the black line is due to the presence of patches of gas moving towards the star, creating a red-shifted component at 20 km/s in the zone (b) as seen in figure \ref{fig:velocities}.
\par
\begin{figure*}[ht!]
%\centerline{\includegraphics[width=\linewidth]{absorbance.png}}

\centerline{\includegraphics[width=0.33\linewidth]{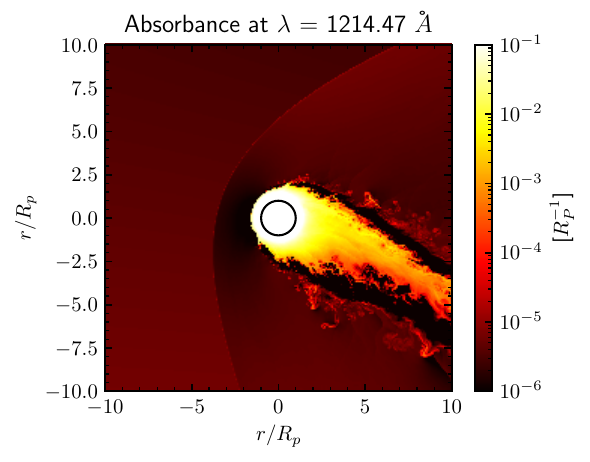}
\includegraphics[width=0.33\linewidth]{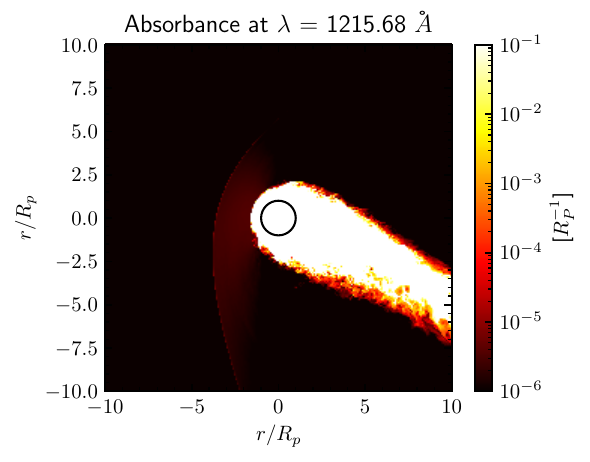}
\includegraphics[width=0.33\linewidth]{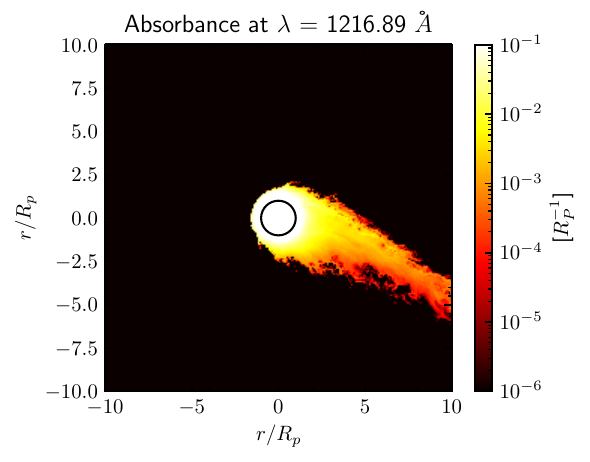}
}
\caption{Instantaneous absorbance coefficient at different wavelengths. \textit{From left to right}: Absorbance for $\lambda$ = 1214.47 {\AA}, $\lambda$ = 1215.68 {\AA} and for $\lambda$ = 1216.89 {\AA}, corresponding to different parts of the Ly-$\alpha$ profile shown in figure \ref{fig:Lymanillustration} (from left to right, the panels correspond to the magenta, orange and cyan vertical lines of Figure \ref{fig:Lymanillustration}). }
\label{fig:dtau}
\end{figure*}

To understand how well the gas absorbs, we calculate the absorbance as follows: $\kappa_{\lambda_k}$ = $n_{H}(x,y)\ \sigma(x,y,\lambda_k)$ (also equal to $d\tau(x,y,\lambda_k)/dx$). It is a property of the gas, independent of the numerical grid adopted.
Figure \ref{fig:dtau} illustrates the absorbance $\kappa_{\lambda_k}$ at different wavelengths represented by the pink, orange, and cyan dashed curves in figure \ref{fig:Lymanillustration}. We observe that the absorbance is higher on the short-wavelength wing of the Ly-$\alpha$ at $\lambda$ = 1214.47 {\AA} compared to the long-wavelength wing at $\lambda$ = 1216.89 {\AA}. Near the centre of the line at $\lambda$ = 1215.68 {\AA}, the gas absorbs on a larger spatial scale, with an additional signal coming from the interface between the planetary and stellar wind.

\subsection{Impact of the planetary mass on the Ly-$\alpha$ profile}

\begin{figure}[h]
\centerline{\includegraphics[width=\linewidth]{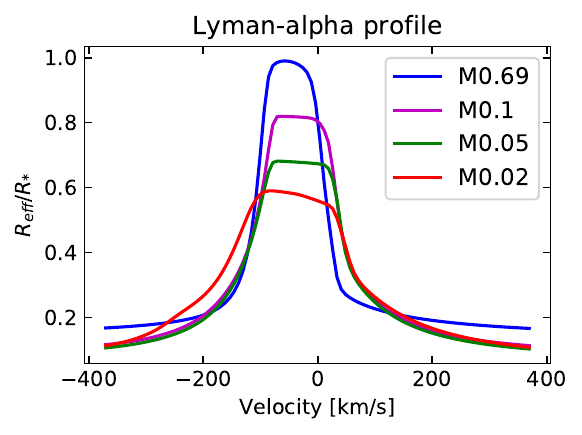}}
\caption{Illustration of the impact of the planetary mass on the Ly-$\alpha$ signal for the different cases normalised with the Sun radius, averaged over 42 hours.}
\label{fig:Lyman_mass}
\end{figure}

The structure of the planetary atmosphere depends on the stellar parameters and on how the deposited energy is used by the planetary wind to overcome the gravitational potential of the planet. Low mass planets have atmospheres that are more extended than more massive planets than higher planet mass planets. This can impact the shape and amplitude of the Ly-$\alpha$ profile. This can impact the shape and amplitude of the Ly-$\alpha$ profile. Figure \ref{fig:Lyman_mass} shows the Ly-$\alpha$ profile (now normalised to $R_\star$) for our four different cases: M0.69, M0.1, M0.05 and M0.02 described in table \ref{table:1}.

As expected, the effective Ly-$\alpha$ radius of the planet normalised to the stellar radius ($R_\star$ = 7 $\times$ 10$^{10}$ cm) is larger for the more massive planet in the core of the line, and goes down to about 0.5 on average for M0.02. Interestingly, all profiles are blue-shifted with similar speeds ranging from 50 km/s in the core of the line and up to 100 km/s on the left wing of the line where it is observable. The lifetime of hydrogen in the planetary tail depends on the rate of photoionisation but also on the dynamics of the winds (planetary + stellar). To observe neutral hydrogen at such high speeds, the role of the stellar wind which accelerates the planetary wind is essential.
\par
We observe asymmetries at the heart of the line which differ from one case to another. The "irregularities" observed in case M0.02 are caused by the presence of patches of neutral gas around the planet. Indeed, the instantaneous shape of the absorption line is dictated by the transient bursts of the planetary wind rather than by the characteristics of the steady-state wind. If the planetary wind is more turbulent, the asymmetry at the heart of the line will increase.

\section{Effect of secondary ionisation on Ly-$\alpha$}\label{sec:5}

Let us now study the impact of secondary ionisation on the Ly-$\alpha$ profile by analysing the two extreme cases of our sample: the cases M0.69 (in blue) and M0.02 (in red). We focus only on the qualitative aspect, and a comparison with observations will be made in a future study.  
\par
\begin{figure}[h!]
\centerline{\includegraphics[width=\linewidth]{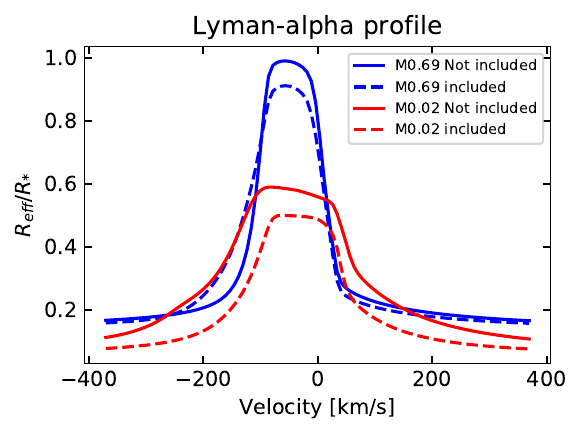}}
\caption{Illustration of the impact of secondary ionisation on the Ly-$\alpha$ profile for cases M0.02 in red and M0.69 in blue. Solid lines represent cases without photoelectrons, while dashed lines represent cases with photoelectrons normalised with the Sun radius. These profiles are also averaged over a period of 42 hours.}
\label{fig:Lyman_photoelectrons}
\end{figure}

We observe that when the secondary ionisation is included (dashed line in the figure \ref{fig:Lyman_photoelectrons}), the amplitude of the Ly-$\alpha$ is generally reduced at positive velocities (those more closely connected with the dayside of the planet). The reason for this reduction is that the size of the neutral cloud is reduced as the ionization proceeds faster.
The same effect seems to occur on the night side of the M0.02 simulation. 
The situation for the M0.69 case is somewhat more complex. Looking into the details of the flowfield, we found that the simulation with the detailed photoelectron simulation exhibits on the nightside a reduced number density of neutrals but higher velocities. The combination of these two features appears as the broadening of the  Ly-$\alpha$ absorption signal at negative velocities in figure \ref{fig:Lyman_photoelectrons}). This serves as a reminder that multiple physical processes can combine differently in ways that are difficult to anticipate.

Both cases M0.02 and M0.69 show a decrease of 0.1 R$_{\rm eff}$/R$_{*}$ around the core of the line. However, on the left wing, the signal is increased by 0.1 R$_{\rm eff}$/R$_{*}$ for M0.69 while decreased by 0.2 for the case M0.02. The effect of photoelectrons is therefore even more important for low-mass planets. We observe an increase of the signal at shorter wavelengths for the case M0.69 when photoelectrons are included, which is the spectral region that is observable. The origin of this interesting feature is not yet fully clear, and a more detailed investigation will be carried out in a future work to assess its significance at these and other planets for which Ly-$\alpha$ observations have been made. Faster ionisation when photoelectrons are included tends to shrink the cloud of neutrals that absorb the Ly-$\alpha$ photons of the star. The change in the mass-loss rate, however, has additional implications on the dynamics of the atmosphere, which may affect the core and the wings of the absorption line differently.

\section{Impact of stellar flares on Ly-$\alpha$}\label{sec:6}

Stellar flares are frequent on young low-mass stars, notably for spectral types M or K around which many detected exoplanets orbit, but also on more massive stars like our Sun of spectral type G \citep{hawley2014kepler}. These events are very energetic, impulsive and difficult to predict, and it is potentially important to take into account their influence on exoplanetary atmospheres when interpreting transit observations.  
\par
There are currently very few studies or modelling of flares and their impact on exoplanetary atmospheres. \citet{hazra2022impact} studied the impact of these flares on HD189733b's atmosphere using 3D radiation hydrodynamic simulations of atmospheric exhaust that include heating by stellar photons using a constant flare model. 
\par
\begin{figure}[h]
\centerline{\includegraphics[width=0.9\linewidth]{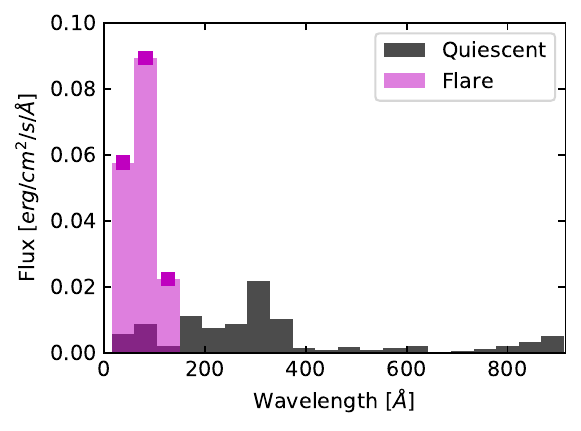}}
\caption{Modelling of the flare on the solar spectrum used in our simulations.}
\label{fig:Flare_Flux}
\end{figure}

In order to know if a dynamic event originating from the star influences the loss of atmospheric mass, we modelled and parameterized the flare as time-varying multiplicative factor $f_{\lambda}(t)$ applied to the X-ray part of the stellar spectrum and defined as follows:

\begin{equation}
\label{eq:flare}
f_{\lambda}(t) = \left\{
    \begin{array}{ll}
        1 & t < t_0  \\
        20 \times e^{(t - t_0)}  & t_0 < t < t_1 \\
        10 \times e^{-2.06\times10^{-4}(t - t_1)}  &  t_1 < t < t_2 \\
        1 & t > t_2
    \end{array}
\right.,
\end{equation}

with t in seconds. It increases rapidly in order to reach at t$_1$ = 300 s an X-ray spectrum 10 times greater, then a slow decrease until returning at t$_2$ = 3 h to its initial state. We only multiplied the incoming stellar flux by $f_{\lambda}$ on the most energetic segments of the solar spectrum $\lambda$ $<$ 100 {\AA} corresponding to the X-ray range (see figure \ref{fig:Flare_Flux}).

\begin{figure}[h]
\centerline{\includegraphics[width=\linewidth]{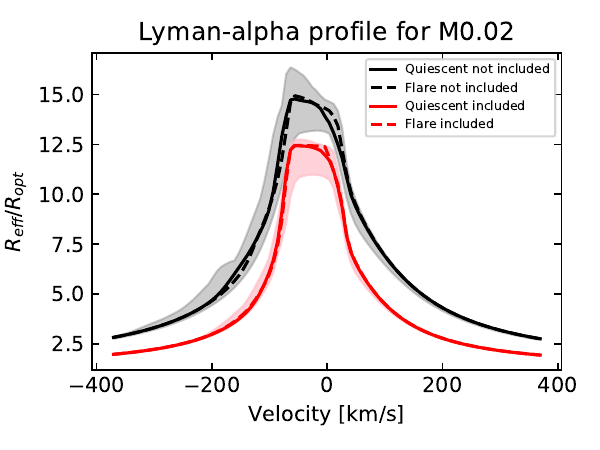}}
\centerline{\includegraphics[width=\linewidth]{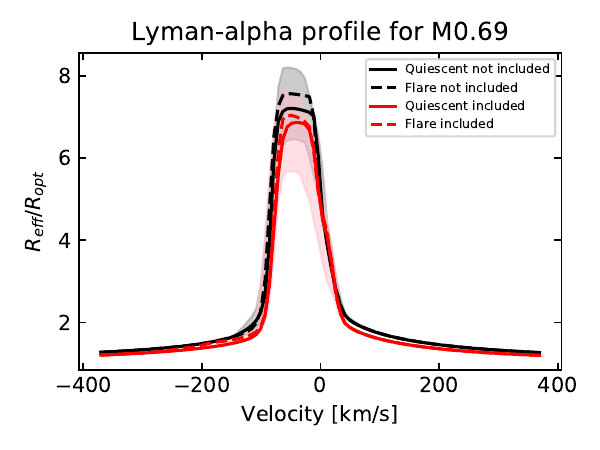}}
\caption{
Illustration of the impact of a flare (dashed line) on the Ly-$\alpha$ profiles of cases M0.02 (upper panel) and M0.69 (lower panel) with (in red) and without (in black) photoelectrons, averaged over 42 hours.}
\label{fig:Flare}
\end{figure}

Let us now look at the impact of the flare on the Ly-$\alpha$ profile in figure \ref{fig:Flare} case M0.02 (upper panel), and on case M0.69 (lower panel). The Ly-$\alpha$ profile at the peak of the flare is modelled by the dashed lines. We observe, in all cases, that the effect of the flare is located within the natural temporal variability of the Ly-$\alpha$ profile (pink and gray zones, temporal average of the profiles over 42 hours) whether secondary ionisation is included or not. 

\par 
This means that we would not be able to directly observe an flare on these profiles. If we calculate the shortest heating timescale (by comparing the total energy $E = P_T /(\gamma -1) + \rho {\bf u}^2/2$ with the energy allocated to heating H, $t_{\rm heat} = {
\rm min}(E/H)$) for the M0.02 case we find t$_{heat}$ = 51.08 h and for the M0.69 case t$_{heat}$ = 9.89 h, during the peak of the flare. The time of the flare is therefore too short compared to the characteristic heating times of the planet to disturb the atmosphere in all cases. Therefore, we do not observe a change in the mass-loss rate. Nevertheless, it is possible that flares of greater intensity, or that a fast repetition of closely flares could temporarily modify the structure of the atmosphere, especially for the most massive planets due to theirs lower heating timescales. Our results differ from  \citep{hazra2022impact}, where they maintain the conditions for a longer period of time in order to let their model reach steady-state. We, however, consider a realistic temporal evolution of the XUV flux of a flare.

\section{Conclusion}\label{sec:7}

After a first study in 1D \citealt{gillet2023self} and its limitations due to the simplified geometry, we have extended our model in a more realistic 2D geometry which integrates both a day part and a night part for the atmosphere. Thanks to this 2D geometry, we could evaluate the impact of photoelectrons on the Ly-$\alpha$ absorption, which is one of the main observables of atmospheric escape. We simulated planets with the same masses as the 1D cases, and studied their effect on heating, ionisation and excitation while comparing them to the 1D results. We also studied the interaction between the stellar wind and the planetary wind.
\par
The planetary wind is formed by the interaction of the host star's XUV radiation with the planet's atmosphere, creating an envelope of partially ionised gas directed toward the ionised stellar wind. The interaction of the two winds strikes a balance between thermal pressure and ram dynamic pressure. We were able to see the classic features of multiscale vortex formation. This interaction forms a shock in front of the planet, whatever the resolution of the computational grid, with temperatures reaching several million kelvins at the level of this interaction zone by compression of the gas.
\par
In the absence of secondary ionisation, the heating rate peaks at similar altitudes in 1D and 2D, but the 2D behaviour is more complex. Photoionisation is primarily affected at higher altitudes in 2D compared to 1D, with significant differences when secondary ionisation is included. Overall, the ionisation rate is stronger in 2D due to a greater amount of neutrals at these altitudes. On the other hand, the lowest altitudes are only slightly impacted by the geometry of our simulations. In 2D, a decrease in mass-loss rates is observed, compared to the equivalent 1D models, with or without photoelectrons. In addition, the inclusion of photoelectrons decreased the mass-loss rate by 22\% to 48\% in the studied sample, similarly as observed in 1D \citet{gillet2023self}, showing the impact of secondary ionisation on low-mass planets. 
\par
In order to link the simulations with the observations, we examined the origin of the Ly-$\alpha$ signal in the atmosphere of case M0.69. We noticed that the signal depends on the density of neutral hydrogen particles, the temperature and the velocity of the gas. The signal contribution is significant near the planet where neutral densities are high, but zero in regions dominated by the ionised stellar wind. The shape of the absorption line is related to the structure of the neutral gas flow. The analysis showed that the largest contributions to the Ly-$\alpha$ signal come from close to the planet. However, the high velocities of the gas towards the observer induce a blue shift of the signal, notably at 100 km/s, which is greater than the velocities observed in 1D and is linked to the interaction with the stellar wind. We do not take into consideration energetic neutral atoms but a future work will add charge exchange between the two winds. 
\par
The structure of an exoplanet's atmosphere depends on the stellar radiative output and on the use of that energy by the atmospheric gas to escape the gravitational potential. In our simulations, low-mass planets have a more extended atmosphere than those with higher masses relative to their radii. This directly affects the Ly-$\alpha$ profile. All profiles are blue-shifted at similar speeds but asymmetries at the core of the line vary from one case to another, they are linked to the dynamics of planetary and stellar winds.
\par
We have discussed the qualitative aspects of how secondary ionisation affects the Ly-$\alpha$ profile. We generally observed a decrease in the amplitude of the Ly-$\alpha$ signal, which is more significant for the low-mass (M0.02) case. As photoelectrons increase the ionisation rate of the atmosphere, thereby reducing the presence of neutrals in the upper atmosphere, we saw that this also directly affected the Ly-$\alpha$ profile.
\par
Stellar flares are common on young, low-mass stars of spectral type M or K, as well as on more massive stars like our Sun of spectral type G. These events, which are difficult to predict and energetic, can affect the atmospheres of transiting exoplanets. We examined the impact of a solar flare on two simulated Jovian and mini-Neptune mass planets orbiting a solar-like star. We find that the effect of the flare lies within the natural variability of the Ly-$\alpha$ profile, even when considering secondary ionisation. The profiles do not show a significant change in mass-loss. However, we believe that more intense or fast and repeated flares could temporarily change the structure of the atmosphere. 
\\

\begin{acknowledgements}
We acknowledge funding from the Programme National de Planétologie (INSU/PNP). A.S. acknowledges funding from the European Union’s Horizon-2020 research and innovation programme (grant agreement no. 776403 ExoplANETS-A) and the PLATO/CNES grant at CEA/IRFU/DAp, the French Agence Nationale de la Recherche (ANR) project STORMGENESIS \#ANR-22-CE31-0013-01, and the European Research Council project ExoMagnets (grant agreement no. 101125367)
\end{acknowledgements}

\bibliographystyle{aa}
\bibliography{sample631}{}

\appendix

\newpage

\end{document}